\documentclass[journal,10pt]{IEEEtran}

\usepackage{cite}
\usepackage{graphicx}
\usepackage{amsmath}
\usepackage{array}
\usepackage{mdwmath}
\usepackage{amssymb}
\usepackage{mdwtab}
\usepackage{stfloats}
\usepackage[tight,footnotesize]{subfigure}
\usepackage{amsmath,amsthm}
\usepackage{threeparttable}
\usepackage{color}
\usepackage{url}
\usepackage{algpseudocode}
\usepackage{algorithm}
\usepackage{multicol}
\usepackage{amssymb}
\usepackage{epstopdf}
\usepackage{flushend}

\algrenewcommand{\algorithmicrequire}{\textbf{Input:}}
\algrenewcommand{\algorithmicensure}{\textbf{Output:}}

\newtheorem{lemma}{Lemma}

\begin{document}

\title{\LARGE Performance of RIS-Assisted Full-Duplex Space Shift Keying With Imperfect Self-Interference Cancellation}

\author{Xusheng Zhu, Wen Chen, \IEEEmembership{Senior Member, IEEE}, Qingqing Wu, \IEEEmembership{Senior Member, IEEE}, Ziwei Liu, and Jun Li, \IEEEmembership{Senior Member, IEEE}
\thanks{
(\emph{Corresponding author: Wen Chen}).}
\thanks{X. Zhu, W. Chen, Q. Wu, and Z. Liu are with the Department of Electronic Engineering, Shanghai Jiao Tong University, Shanghai 200240, China (e-mail: xushengzhu@sjtu.edu.cn; wenchen@sjtu.edu.cn; qingqingwu@sjtu.edu.cn; ziweiliu@sjtu.edu.cn).}
\thanks{J. Li is with the School of Electronic and Optical Engineering, Nanjing University of Science Technology, Nanjing 210094, China (email: jun.li@njust.edu.cn).}
}

\markboth{}
{}
\maketitle
\begin{abstract}
%
In this paper, we consider an full-duplex (FD) space shift keying (SSK) communication system, where information exchange between two users is assisted only by a reconfigurable intelligent surface (RIS). In particular, the impact of loop inter- ference (LI) between the transmit and receive antennas as well as residual self-interference (SI) from the RIS is considered. Based on maximum likelihood detector, we derive the conditional pair- wise error probability and the numerical integration expression for the unconditional pairwise error probability (UPEP). Since it is difficult to find a closed-form solution, we perform accurate estimation by Gauss-Chebyshev quadrature (GCQ) method. To gain more useful insights, we derive an expression for UPEP in the high signal-to-noise ratio region and further give the average bit error probability (ABEP) expression. Monte Carlo simulations were performed to validate the derived results. It is found that SI and LI have severe impacts on system performance. Fortunately, these two disturbances can be well counteracted by increasing the number of RIS units.
\end{abstract}

\section{Introduction}

Spatial modulation (SM) activates a single transmit antenna at each time slot, which exploits both the signal and spatial domains, thus enhancing spectral efficiency \cite{zhu2021performance}.
It is worth noting that SM is presented for multiple-input multiple-output (MIMO) architecture to realize multiplexing gains.
To focus on the spatial domain of SM, \cite{ebase2017index} proposed the space shift keying (SSK) scheme, which is a simplified version of SM by discarding the signal domain part.
In \cite{zhu2023qua}, the quadrature spatial scattering modulation scheme operating in the millimeter wave (mmWave) band was presented by employing hybrid beam techniques.

Reconfigurable intelligent surface (RIS), composed of a large number of low-cost passive units, can significantly improve the signal quality by adjusting the incident and reflected signals\cite{Taosum2021}.
Since each unit of the RIS can be adjusted independently, the quality of the desired signal can be enhanced by tuning the reflection phase shift.
Inspired by this, schemes on RIS-assisted SM and its variants have been extensively studied.
Specifically, \cite{canbilen2020reconfigurable} presented the RIS-assisted SM and SSK systems and provided the detailed theoretical derivation of their average bit error probability (ABEP), in which the RIS is deployed close to the transmitter and the link from the RIS to the receiver is considered.
In practical situations, it is pretty challenging to guarantee that the channel estimation is entirely accurate. Accordingly, \cite{li2021space} investigated the performance of the RIS-SSK system in the presence of channel estimation errors.
It is worth mentioning that when the RIS controller fails to have access to channel status information, there is no way to adjust the reflected phase shift. For this reason, the \cite{bouhlel} investigated the blind RIS-SSK scheme and studied the ABEP performance with transceiver impairments.
Additionally, \cite{zhu2021ris} proposed a novel architecture adapted to mmWave band information transmission through the incorporation of RIS into the spatial scattering modulation system.
However, the previous work fails to address the study of full-duplex (FD) systems. In particular, RIS can also be leveraged to assist FD users in communication \cite{sharma2021inte}.
Although \cite{zhufull} proposed the RIS-FD-SSK scheme, the effect of RIS reflection self-interference (SI) on system performance is not considered.

To this end, in this work, we mathematically study the impact of imperfect self-interference cancellation (SIC) on RIS-FD-SSK systems. Specifically, we simultaneously consider the effects of residual transmitted SI and loop-interference (LI) reflected by RIS on the performance of the RIS-FD-SSK system. Specifically, the maximum likelihood (ML) detection algorithm is used to realize the decoding of the transmitting antenna. Then, the derivation of the analytical ABEP is realized by using the central limit theorem (CLT) approach. In addition, we also employ the Gaussian-Chebyshev quadrature (GCQ) method to achieve accurate estimation of the integration results, where the analytical results are validated by Monte Carlo simulations.
It is shown that when the total number of reflective elements is relatively small, SI and LI have a serious impact on the performance of the system. With the increase of the number of reflective elements, the interference on the RIS-FD-SSK system.

\section{System Model}
\begin{figure}[t]
\centering
\includegraphics[width=7cm]{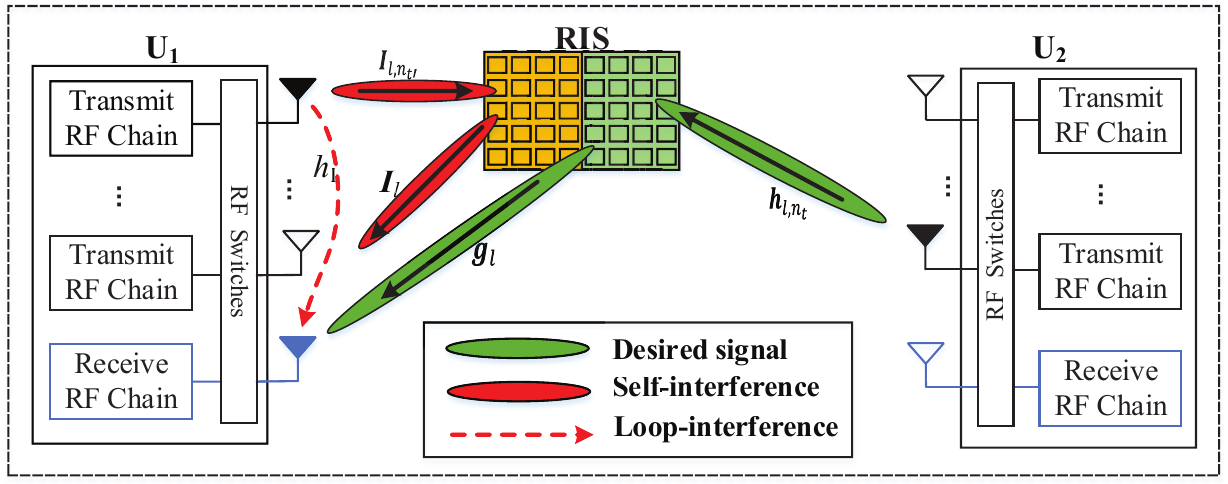}
\caption{\small{System model.}}
\label{Fig1}
\vspace{-10pt}
\end{figure}

As depicted in Fig. \ref{Fig1}, we consider a RIS-FD-SSK system architecture that consists of a RIS, User 1 $(U_1)$, and User 2 $(U_2)$.
Due to severe obstacle obstructions, there is no direct link between $U_1$ and $U_2$.
For this reason, the $2L$-element RIS is deployed in the channel to enable information exchange between $U_1$ and $U_2$,
where the RIS is equally divided into left and right halves to aid communication from $U_1$ to $U_2$ and from $U_2$ to $U_1$, respectively.
Moreover, both $U_1$ and $U_2$ operate in FD mode and are respectively equipped with $(N_t+1)$ antennas,
where one antenna is designated as a receive antenna is shown in blue in Fig. \ref{Fig1}, and the remaining $N_t$ antennas are shown in black as candidate transmit antennas.
Due to the SSK technique, only one transmit antenna in $N_t$ is activated at each transmission time slot.
Meanwhile, the rest of the transmit antennas are in silence. It is worth mentioning that this technique is implemented through radio frequency (RF) switches.
For brevity, we consider the desired signal from $U_2$ via the right half RIS reflection to the $U_1$. It is worth pointing out that a similar approach can be employed to obtain the signal emitted from $U_1$ via RIS reflection to reach the $U_2$ side.
As the SSK technique is not concerned with symbol information, we neglect this part in our work. Then, received signal at the $U_1$ side can be indicated as
\begin{equation}\label{eq1}
{y} \!=\! \sum\nolimits_{l=1}^{L} \eta_{l} e^{j\phi_{l}} h_{l,n_t}  g_{l}  \!+\! h_I  \!+\!\sum\nolimits_{l=1}^L \eta_{l} e^{j\psi_{l}} I_{l}  I_{l,n_{t'}} \! +\! {w}/{\sqrt{P_s}},
\end{equation}
where $\sum_{l=1}^{L} \eta_{l} e^{j\phi_{l}} h_{l,n_t}  g_{l}$ denotes the desired signal transmitted from the $n_t$-th antenna of the $U_2$ via the $L$-element RIS arriving at the receive antenna of the $U_1$ side, where $\eta_l$ and $\phi_l$ means the amplitude and phase shift of $l$-th reflecting element.
Besides, $h_I$ represents the LI signal caused by the transmit antenna to the received signal at the $U_1$ side.
Moreover, {$ \sum\nolimits_{l=1}^L \eta_{l} e^{j\psi_{l}} I_{l}  I_{l,n_{t'}}$} stands for the SI signal that conveys from $n_{t'}$-th antenna of the $U_1$ side via the left half RIS reflection. After that, it is folded back to the receive antenna at the $U_1$ side.
Finally, $w$ denotes complex additive white Gaussian noise with zero-mean and $N_0$ variance, and $P_s$ denotes the transmit power of the signal.
The parameters $h_{l,n_t}$, $g_l$, $I_l$, and $I_{l,n_{t'}}$ denote the fading channel from the $n_t$-th transmit antenna at the $U_2$ side to the $l$-th element of the RIS, the fading channel from the $l$-th element of the RIS to the $U_1$-th receive antenna at the $U_1$ side, the fading channel from the $n_{t'}$-th transmit antenna at the $U_1$ side to the $l$-th element of the RIS, and the fading channel from the $l$-th element of the RIS back to the $U_1$th receive antenna at the $U_1$ side, respectively.

In accordance with this, (\ref{eq1}) can be simplified to
\begin{equation}
\begin{aligned}
{y} \!\!=\!\! \sum\nolimits_{l\!=\!1}^{L} \!\!\alpha_{l,n_t}  \beta_{l}e^{j(\phi_{l}\!-\!\theta_{l,n_t}\!-\!\psi_{l})} \!  +\! h_I \!+\! \!\sum\nolimits_{l=1}^Lf_{l} \! + \!\!\frac{w}{\sqrt{P_s}},
\end{aligned}
\end{equation}
where $h_{l,n_t} = \alpha_{l,n_t}e^{-j\theta_{l,n_t}}$
and $g_l = \beta_{l}e^{-j\psi_l}$.
In particular, $\alpha_{l,n_t}$ and $\beta_{l}$ denote the magnitudes and $\theta_{l,n_t}$ and $\psi_l$ denote the phases of $h_{l,n_t}$ and $g_l$, respectively.
$f_l\sim \mathcal{CN}(0,\kappa (1-{\pi^2}/{16}))$ represents residual SI, where $\kappa\in[0,1]$ indicates the self-interference cancellation (SIC) capability level.
Based on this, the instantaneous signal-to-interference-plus-noise ratio (SINR) at the $U_1$ can be expressed as
\begin{equation}
{\rm SINR} = \frac{|\sum_{l=1}^{L}\alpha_{l,n_t}\beta_l e^{j(\phi_{l}-\theta_{l,n_t}-\psi_l)}|^2}{\Omega_{I}+|\sum\nolimits_{l=1}^Lf_{l}|^2+{N_0}/{P_s} },
\end{equation}
where the $\rm SINR$ value can be maximized by adjusting $\phi_{l} = \theta_{l,n_t}+\psi_l$.
On this basis, the ML detector, which judges the index of the transmitted antenna, can be expressed as

\begin{equation}
{[\hat{n}_t ]}=
\mathop{\mathop{\mathop\mathbf{{\arg\min}}}}_{{n_t \in\{1, \ldots, N_t\}}}
\left|y-\sum\nolimits_{l=1}^{L}\alpha_{l,n_t}\beta_l\right|^{2}.
\end{equation}
\section{Performance Analysis}
In this section, the exact numerical integral expression of ABEP is derived based on the ML detector.
Moreover, high-precision estimation of ABEP expression is evaluated via the GCQ method.
Additionally, asymptotic ABEP expression for the RIS-FD-SSK scheme with imperfect SIC is also provided.
\subsection{Conditional Pairwise Error Probability (CPEP)}
At the $U_1$ side, the CPEP of the RIS-FD-SSK scheme with imperfect SIC can be calculated as
\begin{equation}
\begin{aligned}
P_b =& \Pr\{n_t \to \hat{n}_t|h_{l,n_t},  g_{l}\} \\
= &\Pr\! (|y\! - \!\sum_{l=1}^L \alpha_{l,{n_t}}\beta_l|^2\! >\!|y\!-\!\sum_{l=1}^L \alpha_{l,{\hat{n}_t}}\beta_l e^{-j(\theta_{l,n_t}-\theta_{l,{\hat{n}_t}})}|^2)\\
\overset{(a)}{=}  &\Pr \left(|G_{n_t}|^2-|G_{{\hat{n}}_t}|^2-2\Re\left\{ y (G_{n_t}-G_{{\hat{n}}_t})\right\}\right)\\
=&\Pr (-|G_{n_t}-G_{{\hat{n}}_t}|^2
-2\Re\{ (h_I +\sum\nolimits_{l=1}^Lf_{l}  \\&+ {w}/{\sqrt{P_s}}) (G_{n_t}\!-\!G_{{\hat{n}}_t})\}\!>\!0)
= \Pr(G>0),
\end{aligned}
\end{equation}
where $(a)$ stands for $G_{n_t} = \sum\nolimits_{l=1}^L \alpha_{l,{n_t}}\beta_l$ and $G_{\hat{n}_t} = \sum\nolimits_{l=1}^L \alpha_{l,{\hat{n}_t}}\beta_l e^{-j(\theta_{l,n_t}-\theta_{l,{\hat{n}_t}})}$;
$\Re\{\cdot\}$ denotes the real part of a complex variable.
Note that $G=-|G_{n_t}-G_{{\hat{n}}_t}|^2
-2\Re\{ (h_I + \sum\nolimits_{l=1}^L  f_{l}    + \frac{w}{\sqrt{P_s}}) (G_{n_t}-\!G_{{\hat{n}}_t})\}$ follows Gaussian distribution $\mathcal {N}(\mu_G,\sigma_G^2)$ with
$\mu_G = -|G_{n_t}-G_{{\hat{n}}_t}|^2$ and
$\sigma_G^2 = 2 ({\Omega_{I}+|\sum\nolimits_{l=1}^L  f_{l}  |^2+\frac{1}{\rho} }) |G_{n_t}-G_{{\hat{n}}_t}|^2$.

\begin{equation}\label{cpep}
P_b \!=\!Q\left(\!\frac{-\mu_G}{\sigma_G}\!\right) \!=\!Q\left(\!\sqrt{\frac{|G_{n_t}-G_{\hat{n}_t}|^2}{2({\Omega_{I}\!+|\sum\nolimits_{l=1}^L  f_{l}  |^2\!\!+\!{1}/{\rho} })}}\!\right),
\end{equation}
where $\rho = P_s/N_0$ denotes the average signal-to-noise ratio (SNR).

\subsection{Unconditional Pairwise Error Probability (UPEP)}
It can be observed from (\ref{cpep}) that there exist both variables on the numerator and denominator, and directly bringing in the calculation to solve would greatly increase the computational difficulty. To address this problem, we first solve for the denominator.
According to CLT, we have $E[\sum\nolimits_{l=1}^L  f_{l}  ]=\kappa^2L(1-\frac{\pi^2}{16})$.
In this manner, the CPEP can be simplified as
\begin{equation}
P_b = Q\left(\sqrt{\frac{|R|^2}{2({\Omega_{I}+\kappa^2L(1-\frac{\pi^2}{16})+\frac{1}{\rho} })}}\right),
\end{equation}
where $R=\sum\nolimits_{l=1}^L \alpha_{l,{n_t}}(\beta_l-\beta_l e^{-j(\theta_{l,n_t}-\theta_{l,{\hat{n}_t}})})$.
Since the number of cells in RIS is much larger than one, we adopt CLT to treat $R$ as a Gaussian distribution obeying an expectation of $\mu_R$ with a variance of $\sigma_R^2$.

\begin{lemma}
The expectation and variance of $R$ can be denoted respectively as
\begin{equation}
\begin{aligned}
\mu_R = {(L\pi)}/{4}, \ \ \sigma_R^2 = (L(32-\pi^2))/{16}.
\end{aligned}
\end{equation}
Proof:
To simplify the representation,
we let $\beta=\beta_l$, $\alpha = \alpha_{l,n_t} - \alpha_{l,\hat{n}_t}e^{-j(\theta_{l,\hat{n}_t}-\theta_{l,n_t})}$,
$x = \theta_{l,n_t}, y = \theta_{l,\hat{n}_t}$, and $z = \theta_{l,\hat{n}_t}-\theta_{l,n_t}$, where $\beta\in[-\pi,\pi]$ and $\alpha\in[-\pi,\pi]$ follow independent uniform random variables.
In particular, the CDF of $z$ can be expressed as
\begin{equation}\label{CDfz1}
\begin{aligned}
F_Z(z)&=\Pr\left(Z\leq z\right)=\Pr\left(X-Y\leq z\right)=\Pr\left(X\leq Y+z\right)\\
&=\int_{-\infty}^\infty\int_{-\infty}^{y+z}f(x,y)dxdy.
\end{aligned}
\end{equation}
Depending on (\ref{CDfz1}), we can use $f_Z(z)=\frac{d}{dz}F_Z(z)$ to get
$
f_Z(z)=\int_{-\infty}^\infty f(y, y+z)dy$.
Since the random variables $X$ and $Y$ are independent of each other, we obtain $f_Z(z)=\int_{-\infty}^\infty f(y)f(y+z)dy$.
Then, we partition the variable of $y$ into four intervals to address each of them as follows:
{\bf Case 1:} For $z\leq -2\pi$, we have $f_Z(z)=0$.
{\bf Case 2:} For $-\pi\leq -z+\pi\leq \pi$, i.e., $0\leq z\leq 2\pi$, we get $f_Z(z)=\int_{-\pi}^{-z+\pi}f(y)f(y+z)dy=\frac{2\pi-z}{4\pi^2}$.
{\bf Case 3:} For $-\pi\leq -z-\pi\leq \pi$, i.e., $-2\pi\leq z\leq 0$, we have $f_Z(z)=\int_{-z-\pi}^{\pi}f(y)f(y+z)dy=\frac{2\pi+z}{4\pi^2}$.
{\bf Case 4:} For $z \geq 2\pi$, we get $f_Z(z)=0$.
With this in mind, we have
\begin{equation}\label{mevai}
\begin{aligned}
E[\alpha_{l,\hat{n}_t}e^{jz}]&=E[\alpha_{l,\hat{n}_t}\cos{z}]+E[\alpha_{l,\hat{n}_t}\sin{z}],\\
Var[\alpha_{l,\hat{n}_t}e^{jz}]&=Var[\alpha_{l,\hat{n}_t}\cos{z}]+Var[\alpha_{l,\hat{n}_t}\sin{z}],
\end{aligned}
\end{equation}
where $E[\alpha_{l,\hat{n}_t}]=\frac{\sqrt{\pi}}{2}$ and $Var[\alpha_{l,\hat{n}_t}]=\frac{4-\pi}{4}$. Next, we aim at obtaining the mean and variance of $\cos z$ and $\sin z$, the mean of $\cos z$ and $\sin z$ can be respectively expressed as
\begin{equation}
\begin{aligned}
E[\cos z]\!=\!\int_{-2\pi}^0\!\!\frac{2\pi\!+\!z}{4\pi^2}\cos zdz\!+\!\!\int_0^{2\pi}\!\!\frac{2\pi\!-\!z}{4\pi^2}\cos zdz
\!=\!0,\\
E[\sin z]\!=\!\int_{-2\pi}^0\!\!\frac{2\pi\!+\!z}{4\pi^2}\sin zdz\!+\!\!\int_0^{2\pi}\!\!\frac{2\pi\!-\!z}{4\pi^2}\sin zdz
\!=\!0.
\end{aligned}
\end{equation}
\begin{figure}[t]
  \centering
  \includegraphics[width=6cm]{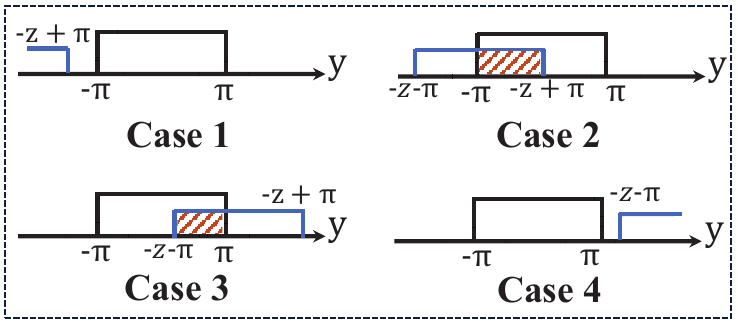}\\
 \caption{\small{The integration interval of the variable $y$ in different cases.}}\label{case}
\end{figure}

Resort to $E[XY]=E[X]E[Y]$, we have
$E[\alpha_{l,\hat{n}_t}\cos{z}]=0$ and $E[\alpha_{l,\hat{n}_t}\sin{z}]=0$.
To obtain the variance, $E[\cos^2 z]$ and $E[\sin^2 z]$ can be respectively derived as
\begin{equation}
\begin{aligned}
&E[\cos^2 z]=\int_{-2\pi}^0\!\!\frac{2\pi+z}{4\pi^2}\cos^2 zdz+\int_0^{2\pi}\!\!\frac{2\pi-z}{4\pi^2}\cos^2 zdz\\
&=\int_{-2\pi}^0\!\!\!\! \frac{(2\pi\!+\!z)(1\!+\!\cos 2z)}{8\pi^2}dz+\!\!\int_0^{2\pi}\!\!\frac{(2\pi\!-\!z)(1\!+\!\cos 2z)}{8\pi^2}dz\\
&=\frac{1}{8\pi^2}\left[\int_{-2\pi}^0(2\pi+z)dz+\int_0^{2\pi}(2\pi-z)dz\right]=\frac{1}{2},
\end{aligned}
\end{equation}
\begin{equation}
\begin{aligned}
&E[\sin^2 z]=\int_{-2\pi}^0\!\!\frac{(2\pi+z)\sin^2 z}{4\pi^2}dz+\int_0^{2\pi}\!\!\frac{(2\pi-z)\sin^2 z}{4\pi^2}dz
\\
&=\int_{-2\pi}^0\!\!\!\frac{(2\pi\!+\!z)(1\!-\!\cos 2z)}{8\pi^2}dz+\!\!\int_0^{2\pi}\!\frac{(2\pi\!-\!z)(1\!-\!\cos 2z)}{8\pi^2}dz\\
&=\frac{1}{8\pi^2}\left[\int_{-2\pi}^0(2\pi+z)dz+\int_0^{2\pi}(2\pi-z)dz\right]=\frac{1}{2}.
\end{aligned}
\end{equation}
Subsequently, the variance of $\sin z$ and $\cos z$ can be respectively obtained as $Var[\sin z]=E[\sin^2z]-E[\sin z]=1/{2}$ and $Var[\cos z]=E[\cos^2z]-E[\cos z]={1}/{2}$.
Based on the relationship between the expectation and variance of multiplying two independent random variables, i.e., $Var[XY]=E^2[X]Var[Y]+Var[X]E^2[Y]+Var[X]Var[Y]$, we have $E[\alpha_{l,\hat{n}_t}\cos{z}]=0$, $Var[\alpha_{l,\hat{n}_t}\cos{z}]={1}{2}$,
$E[\alpha_{l,\hat{n}_t}\sin{z}]=0$, and $Var[\alpha_{l,\hat{n}_t}\sin{z}]={1}/{2}$.
From this, the expectation and variance of the error detection symbols with RIS phase shift can be represented as
$E[\alpha_{l,\hat{n}_t}e^{jz}]=0$ and $Var[\alpha_{l,\hat{n}_t}e^{jz}]=1$.
Utilizing the addition rule for random variables, we have
\begin{equation}
E[\alpha] = {\sqrt{\pi}}/{2}, \ \ \ Var[\alpha] = ({8-\pi})/{4}.
\end{equation}
Afterward, the corresponding proof is completed by exploiting the rule of multiplication of independent random variables $\alpha$ and $\beta$.
$\hfill\blacksquare$
\end{lemma}
Without loss of generality, let us define $\gamma = |R|^2$.
Herein, the UPEP can be given by
\begin{equation}\label{upep1}
\overline P_b = \int_{0}^{\infty}Q\left(\sqrt{\frac{\gamma}{2({\Omega_{I}+\kappa^2L(1\!\!-\!\!\frac{\pi^2}{16})+\frac{1}{\rho} })}}\right)f(\gamma)d\gamma.
\end{equation}
Substituting $Q(\gamma)=\frac{1}{\pi}\int_0^{\frac{\pi}{2}}\exp\left(-\frac{\gamma^2}{2\sin^2\vartheta}\right)d\vartheta$ into (\ref{upep1}) \cite{zhu2023qua}, the UPEP can be equivalently formulated as
\begin{equation}\label{pasfwef}
\overline P_b \!=\!\frac{1}{\pi} \!\! \int_{0}^{\infty}\!\!\int_0^{\frac{\pi}{2}}\!\!\exp\left(\!\frac{-\gamma}{4({\Omega_{I}\!\!+\!\!\kappa^2L(1\!\!-\!\!\frac{\pi^2}{16})\!\!+\!\!\frac{1}{\rho} })\sin^2\vartheta}\!\right) \!f(\gamma)d\vartheta\! d\gamma.
\end{equation}
It is quite complicated to tackle (\ref{pasfwef}) directly. For this purpose, by exchanging the order of integration of $\vartheta$ and $\gamma$, we obtain
\begin{equation}\label{upepe3}
\overline P_b \!=\!\frac{1}{\pi} \! \int_0^{\frac{\pi}{2}}\!\!\int_{0}^{\infty}\!\!\exp\left(\!\frac{-\gamma}{4({\Omega_{I}\!\!+\!\!\kappa^2L(1\!\!-\!\!\frac{\pi^2}{16})\!\!+\!\!\frac{1}{\rho} })\sin^2\vartheta}\!\right) \!f(\gamma)d\gamma\!d\vartheta .
\end{equation}
Since $\gamma$ is a non-central chi-square distribution with one degree of freedom, the MGF of $\gamma$ can be characterized as $M_\gamma (x)=\sqrt{\frac{1}{1-2\sigma_R^2 x}}\exp\left(\frac{\mu_R^2x}{1-2\sigma_R^2x}\right)$ \cite{zhufull}
At this point, the (\ref{upepe3}) can be updated as
\begin{equation}\label{pbba1}
\overline P_b =\frac{1}{\pi} \int_0^{\frac{\pi}{2}} \sqrt{\frac{2\varsigma\sin^2\vartheta}{2\varsigma\sin^2\vartheta+\sigma_R^2 \gamma}}\exp\left(\frac{-\mu_R^2\gamma}{4\varsigma\sin^2\vartheta+2\sigma_R^2\gamma}\right) d\vartheta,
\end{equation}
where $\varsigma={\Omega_{I}\!+\!\kappa^2L(1\!-\!\frac{\pi^2}{16})\!+\!\frac{1}{\rho} }$.
To facilitate subsequent analysis, we make
$\vartheta=\frac{\pi}{4}\varphi+\frac{\pi}{4}$.
When $\vartheta=0$ and $\vartheta=\frac{\pi}{2}$, we get $\varphi=-1$ and $\varphi=1$, respectively.
In this respect, the (\ref{pbba1}) can be recast as
\begin{small}
\begin{equation}\label{gcqpre1}
\begin{aligned}
\overline P_b =&\frac{1}{4} \int_{-1}^1 \sqrt{\frac{2\varsigma\sin^2\left(\frac{\pi}{4}\varphi+\frac{\pi}{4}\right)}{2\varsigma\sin^2\left(\frac{\pi}{4}\varphi+\frac{\pi}{4}\right)+\sigma_R^2 \gamma}}\\
&\times\exp\left(\frac{-\mu_R^2\gamma}{4\varsigma\sin^2\left(\frac{\pi}{4}\varphi+\frac{\pi}{4}\right)+2\sigma_R^2\gamma}\right) d\varphi.
\end{aligned}
\end{equation}
\end{small}
It is difficult to obtain closed-form expressions of (\ref{gcqpre1}). As a consequence, we employ the GCQ method to address it as
\begin{equation}\label{upeprg}
\begin{aligned}
\overline P_b =&\frac{\pi}{4G} \sum_{g=1}^G \sqrt{\frac{2(1-\varphi_g^2)\varsigma\sin^2\left(\frac{\pi}{4}\varphi_g+\frac{\pi}{4}\right)}{2\varsigma\sin^2\left(\frac{\pi}{4}\varphi_g+\frac{\pi}{4}\right)+\sigma_R^2 \gamma}}\\
&\times\exp\left(\frac{-\mu_R^2\gamma}{4\varsigma\sin^2\left(\frac{\pi}{4}\varphi_g+\frac{\pi}{4}\right)+2\sigma_R^2\gamma}\right) +R_G,
\end{aligned}
\end{equation}
where $\varphi_g = \cos\left(\frac{2g-1}{G}\pi\right)$, $G$ is the complexity-accuracy trade-off factor, and $R_G$ is the error term that can be ignored due to the high value of $G$.

\subsection{Asymptotic UPEP}
To more intuitively understand the impact of key parameters on the RIS-FD-SSK scheme under imperfect SIC, we require the derivation of asymptotic UPEP expression.
Particularly, taking limits on $\rho$ values, we have $\lim\limits_{\rho\to\infty}\varsigma=\Omega_I+\kappa^2L-({\pi^2\kappa^2L})/{16}$.
After that, we substitute this into (\ref{upeprg}) and disregard the error term. At this time, the asymptotic UPEP can be obtained as
\begin{small}
\begin{equation}\label{asy}
\begin{aligned}
&\overline P_b =\frac{\pi}{4G} \sum_{g=1}^G \sqrt{\frac{2(1-\varphi_g^2)\left(\Omega_I+\kappa^2-\frac{\pi^2\kappa^2}{16}\right)\sin^2\left(\frac{\pi}{4}\varphi_g+\frac{\pi}{4}\right)}{2\left(\Omega_I+\kappa^2-\frac{\pi^2\kappa^2}{16}\right)\sin^2\left(\frac{\pi}{4}\varphi_g+\frac{\pi}{4}\right)+\sigma_R^2 \gamma}}\\
&\times\exp\left(\frac{-\mu_R^2\gamma}{4\left(\Omega_I+\kappa^2-\frac{\pi^2\kappa^2}{16}\right)\sin^2\left(\frac{\pi}{4}\varphi_g+\frac{\pi}{4}\right)+2\sigma_R^2\gamma}\right).
\end{aligned}
\end{equation}
\end{small}

\subsection{ ABEP}
Considering the assumption of uniform error probability, the tight union upper bound of ABEP can be formulated as \cite{li2021space}
\begin{equation}\label{ABEP}
{\rm{ABEP}} \leq \sum\nolimits_{n_t=1}^{N_t}\sum\nolimits_{\hat n_t=1, \hat n_t \neq n_t}^{N_t}\overline P_b N(n_t\to \hat n_t),
\end{equation}
where $N(n_t\to \hat n_t)$ denotes the number of bits in error between the $n_t$ and $\hat n_t$ and $\overline P_b$ represent the error probability.
It is worth mentioning that when $N_t=2$, (\ref{ABEP}) is equal to the exact ABEP value. However, when $N_t>2$, (\ref{ABEP}) represents the union upper bound of ABEP.
\subsection{Convergence and Accuracy Analysis of GCQ Method}
\begin{figure}[t]
\centering
\begin{minipage}[t]{0.45\linewidth}
\centering
\includegraphics[width=4.0cm]{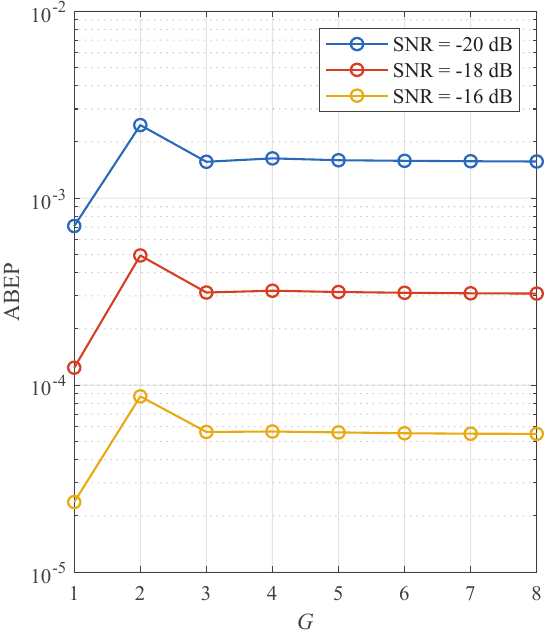}
\caption{\small{Convergence Analysis of GCQ Method.}}
\label{cover}
\end{minipage}
\centering
\begin{minipage}[t]{0.45\linewidth}
\centering
\includegraphics[width=4.0cm]{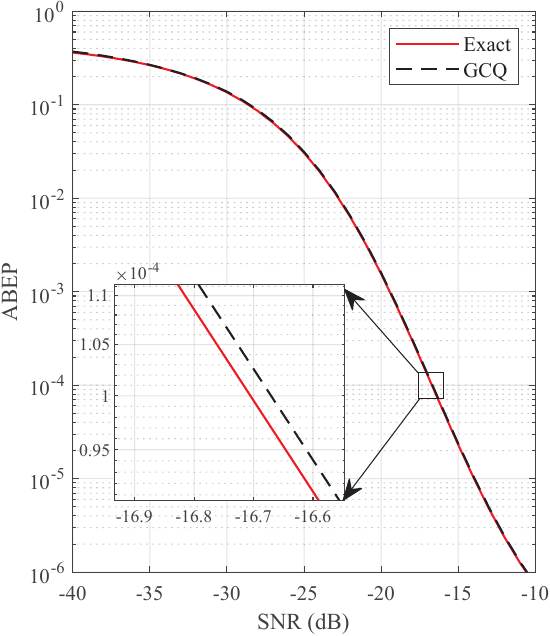}
\caption{\small{Accuracy Analysis of GCQ Method ($G=5$).}}
\label{accau}
\end{minipage}
\end{figure}

To evaluate the ABEP obtained by the GCQ method more intuitively, we plot Fig. \ref{cover} and Fig. \ref{accau} for the analysis of the convergence and accuracy of the GCQ method, respectively.
Note that the number of reflective elements of the RIS is configured to 64 in Figs. 3 and 4.
Specifically, in Fig. \ref{cover}, we observed that the ABEP value obtained by using the GCQ method converges very quickly, and when $G \geq 3$, it reaches a stable state.
Afterward, we provide an evaluation of its accuracy in Fig. \ref{accau}, where the parameter $G=5$.
It is clear that the results obtained with the GCQ method almost overlap with the results achieved with the numerical integration (\ref{gcqpre1}). As can be seen in the enlarged subplot, there is still a gap between the two, and the reason for the gap is caused by the error term $R_G$. The gap can be reduced by increasing $G$.

\begin{figure}[t]
\centering
\begin{minipage}[t]{0.45\linewidth}
\centering
\includegraphics[width=3.8cm]{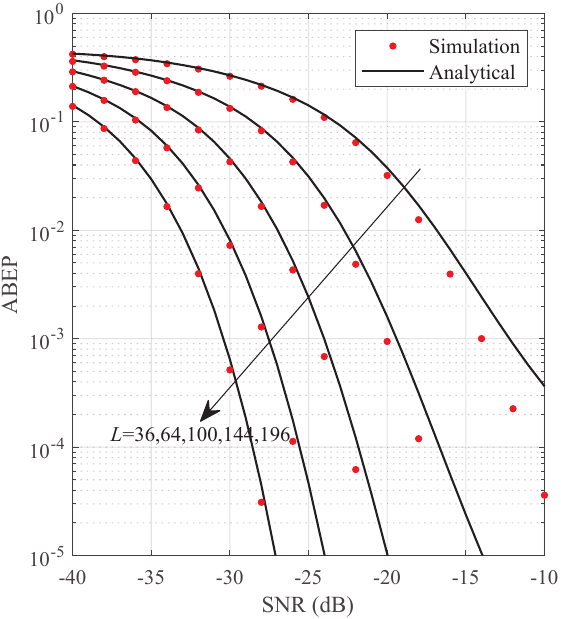}
\caption{\small{Verification of analytical results.}}
\label{Figclt}
\end{minipage}
\centering
\begin{minipage}[t]{0.45\linewidth}
\centering
\includegraphics[width=3.8cm]{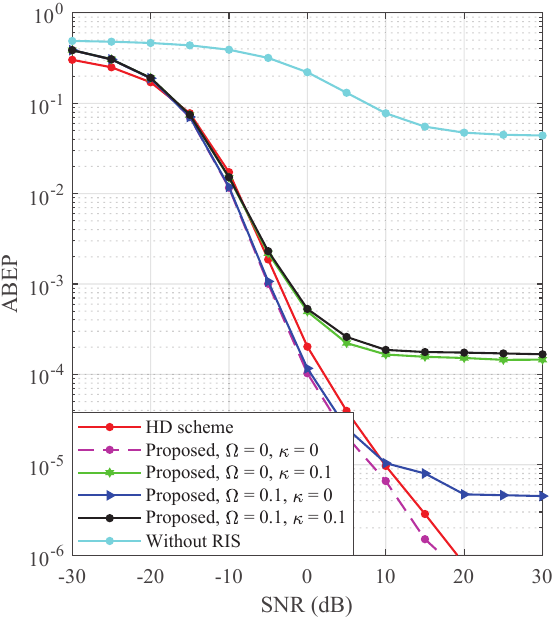}
\caption{\small{Compare with benchmarks.}}
\label{vs}
\end{minipage}
\end{figure}
\section{Simulation Results}
In this section, we investigate the simulation results and analysis curves of ABEP for the RIS-FD-SSK scheme with imperfect SIC. For each simulated value, $10^7$ channels are randomly generated.
Besides, we draw the relationship between the received SNR and ABEP.
Note that the path loss is implicitly incorporated in the receive SNR and therefore is not directly represented in the expression.
Unless otherwise stated, the transmit and receive antennas are defaulted to 2 and 1, respectively.

In Fig. \ref{Figclt}, we validate the theoretical resolution results by using Monte Carlo simulations, where the analytical ABEP is obtained through CLT.
It can be observed from Fig. \ref{Figclt} that the gap between the simulation and analytical results is larger when the number of RISs is 36 and 64. However, as $L$ increases, the gap between them gradually decreases, which means that when the number of RISs is larger, the results obtained with CLT are more accurate.
Consequently, in the subsequent analysis, we adopt the simulation results for illustration if the number of reflecting units of RIS is less than 100, and we adopt the analytical approach to solve for it if the number of reflecting units of RIS is not less than 100.

In Fig. \ref{vs}, we depict the ABEP of the RIS-FD-SSK scheme under the imperfect SIC with respect to its HD and without RIS counterparts.
To make the comparison fair, we unify the transmission rate of the system to 4 bits per channel use.
As expected, the system performs worst in the absence of RIS. This is because each unit of the RIS can be adjusted independently can enhance the transmission quality of the signal employing spatial diversity.
Meanwhile, when neither LI nor SI exists, the ABEP performance of the FD scheme is better than that of the HD scheme. However, when LI or SI is present, the performance ratio of HD's scheme ABEP becomes better with increasing SNR. This is because HD has no SI and HI, and the performance of ABEP is only limited by SNR. In the FD scheme, when the SNR reaches a certain level, SI and LI become the main factors restricting ABEP.

\begin{figure}[t]
 \centering
 \subfigure[$L = 16$]
 {
  \begin{minipage}[b]{0.22\textwidth}
   \centering
   \includegraphics[width=3.8cm]{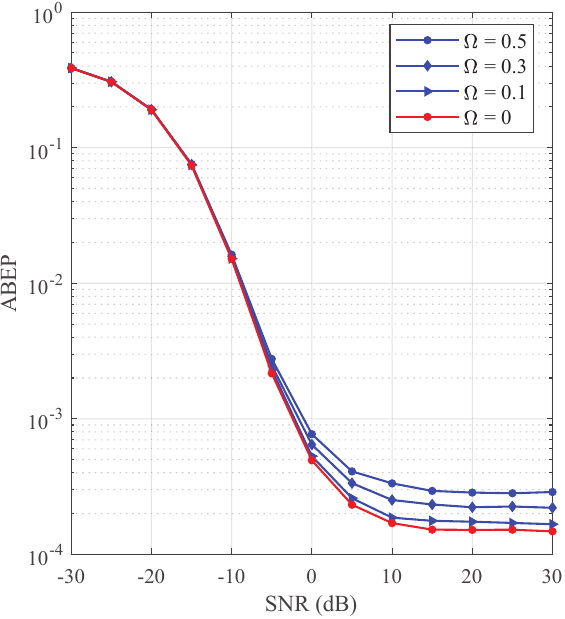}
  \end{minipage}
 }
 \subfigure[$L = 100$]
    {
     \begin{minipage}[b]{0.22\textwidth}
      \centering
      \includegraphics[width=3.8cm]{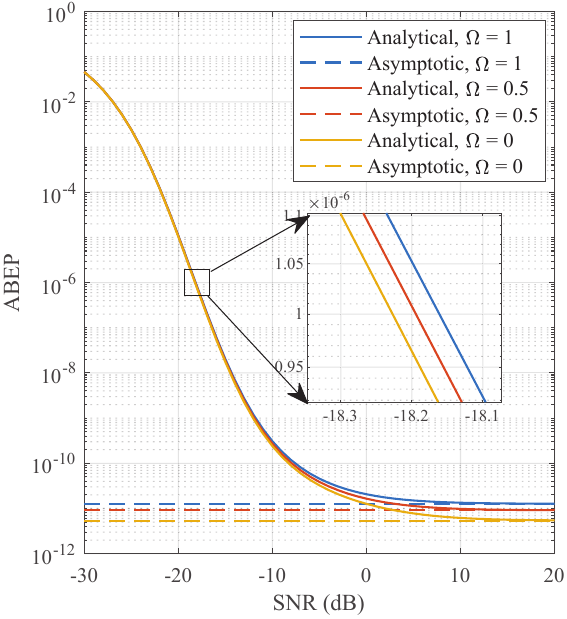}
     \end{minipage}
    }
\caption{\small{Impact of LI on the performance of RIS-FD-SSK system.}}
\label{figLI}
\end{figure}
\begin{figure}[t]
 \centering
 \subfigure[$L = 16$]
 {
  \begin{minipage}[b]{0.22\textwidth}
   \centering
   \includegraphics[width=3.8cm]{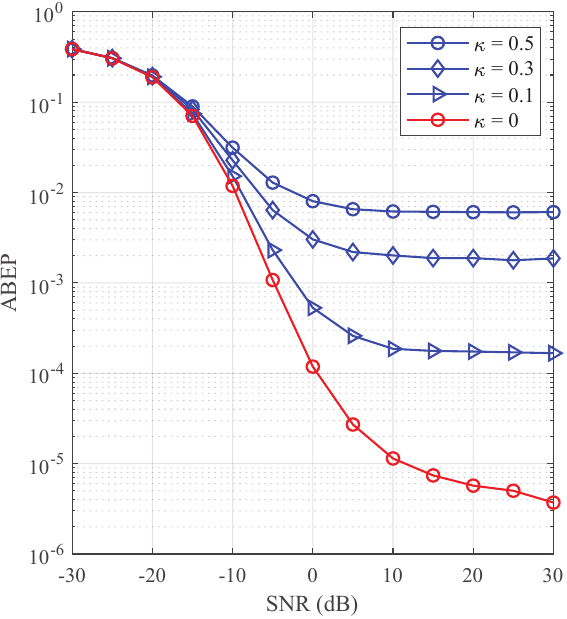}
  \end{minipage}
 }
 \subfigure[$L = 100$]
    {
     \begin{minipage}[b]{0.22\textwidth}
      \centering
      \includegraphics[width=3.8cm]{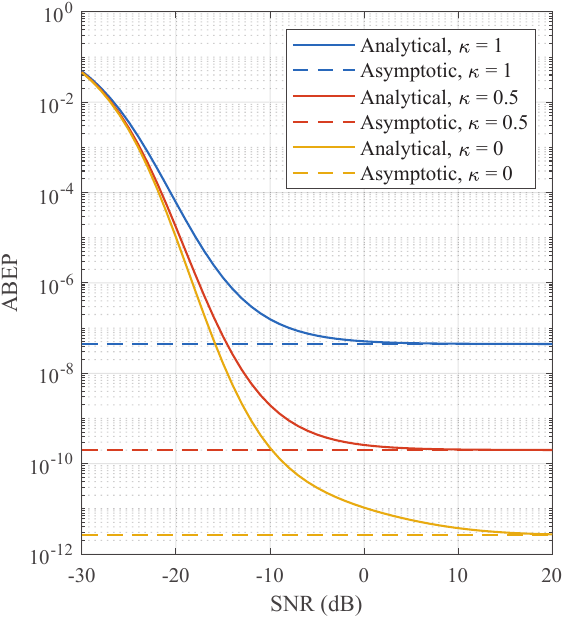}
     \end{minipage}
    }
\caption{\small{Impact of imperfect SIC on RIS-FD-SSK system.}}
\label{figsic}
\end{figure}

In Fig. \ref{figLI}, we depict the impact of LI on the ABEP performance of the system for the RIS-FD-SSK scheme under imperfect SIC, where we set the parameter $\kappa=0.1$.
When $L = 16$ in Fig. \ref{figLI}(a), we can observe that as the value of $\Omega$ becomes smaller, it indicates that the influence of LI is less and thus the performance is better. Moreover, to clarify the effect of LI on the system, we plot a limit state where $\Omega = 0$ indicates the absence of LI interference.
On the other hand, when $L=100$ in Fig. \ref{figLI}(b), we graph the analytical and asymptotic ABEP curves depending on (\ref{upeprg}) and (\ref{asy}), respectively.
It can be observed that the asymptotic and analytical values overlap in the high SNR region, which verifies the asymptotic ABEP derivation. In addition, when ABEP = $10^{-6}$, LI has little effect on the system performance, which indicates that increasing $L$ can improve the noise immunity of the system.

Fig. \ref{figsic} shows the ABEP performance of the RIS-FD-SSK scheme with different SIC levels, where we let the parameter $\Omega$ be set to 0.1.
To be specific, if the RIS has 16 reflective units, the system performance of the proposed RIS-FD-SSK scheme improves as the SIC capability $\kappa$ increases in Fig. \ref{figsic}(a). Compared with LI, the RIS-FD-SSK scheme is more sensitive to SI changes, which is caused by the aggregation and enhancement of interference signals by each unit of RIS.
For the case of $L=100$, the asymptotic ABEP curves match the theoretical results in the SNR region in Fig. \ref{figsic}(b), which further validates the accuracy of the asymptotic results.
At ABEP = $10^{-6}$, the variation of Fig. \ref{figsic}(b) with respect to Fig. \ref{figsic}(a) is not as sensitive, this is because as the number of RIS is larger, the energy intensity of the desired signal is larger and the signal quality is higher.

\section{Conclusion}

In this paper, we investigated the reliability of the RIS-assisted SSK communication system for FD users, where impacts of residual LI and SIC levels are considered.
Based on the ML detector, we derive the integral expression of ABEP via the CLT.
To gain more useful insights, we obtain its convergence and accuracy by employing the GCQ method, where the correctness of the analytically derived results is verified by Monte Carlo simulation.
The results show that when the number of reflection units of RIS is relatively small, the SI and LI can have a detrimental effect on the performance of the system.
However, as the number of RIS elements increases, the effect of interference on RIS-FD-SSK becomes decreasingly smaller.



\begin{thebibliography}{99}

\bibitem{zhu2021performance}
X. Zhu, L. Yuan, Q. Li, Q. Li, L. Jin, and J. Zhang, ``On the performance of 3-D spatial modulation over measured indoor channels," \emph{IEEE Trans. Veh. Technol.}, vol. 71, no. 2, pp. 2110-2115, Feb. 2022.

\bibitem{ebase2017index}
J. Jeganathan, A. Ghrayeb, L. Szczecinski, and A. Ceron, ``Space shift keying modulation for MIMO channels,"  \emph{IEEE Trans. Wireless Commun.}, vol. 8, no. 7, pp. 3692-3703, Jul. 2009.

\bibitem{zhu2023qua}
X. Zhu, W. Chen, Z. Li, Q. Wu, and J. Li, ``Quadrature spatial scattering modulation for mmWave transmission," \emph{IEEE Commun. Lett.}, vol. 27, no. 5, pp. 1462-1466, May 2023.

\bibitem{Taosum2021}
A. Ihsan, W. Chen, M. Asif, W. U. Khan, Q. Wu, and J. Li, ``Energy-efficient IRS-aided NOMA beamforming for 6G wireless communications," \emph{IEEE Trans. Green Commun. Netw.}, vol. 6, no. 4, pp. 1945-1956, Dec. 2022.

\bibitem{canbilen2020reconfigurable}
E. Basar, ``Reconfigurable intelligent surface-based index modulation: A new beyond MIMO paradigm for 6G,"  \emph{IEEE Trans. Commun.}, vol. 68, no. 5, pp. 3187-3196, May 2020.

\bibitem{li2021space}
X. Zhu, W. Chen, Q. Wu, and L. Wang, ``Performance analysis of RIS-aided space shift keying with channel estimation errors," [Online] Available: https://arxiv.org/abs/2307.01994.

\bibitem{bouhlel}
A. Bouhlel, M. M. Alsmadi, E. Saleh, S. Ikki, and A. Sakly, ``Performance analysis of RIS-SSK in the presence of hardware impairments," in \emph{ Proc. IEEE 32nd Annu. Int. Symp. Pers., Indoor and Mobile Radio Commun. (PIMRC)}, Helsinki, Finland, 2021, pp. 537-542


\bibitem{zhu2021ris}
X. Zhu, L. Yuan, K. J. Kim, Q. Li, and J. Zhang, ``Reconfigurable intelligent surface-assisted spatial scattering modulation," \emph{IEEE Commun. Lett.},  vol. 26, no. 1, pp. 192-196,
Jan. 2022.

\bibitem{sharma2021inte}
B. C. Nguyen, T. M. Hoang, L. T. Dung, and T. Kim, ``On Performance of two-way full-duplex communication system with reconfigurable intelligent surface," \emph{IEEE Access}, vol. 9, pp. 81274-81285, Jun. 2021.


\bibitem{zhufull}
X. Zhu {et al.} ``RIS-assisted full-duplex space shift keying: System scheme and performance analysis," \emph{IEEE Trans. Green Commun. Netw.}, early access, doi: 10.1109/TGCN.2023.3293913.


\end{thebibliography}
\end{document}